\documentclass[aps,pra,twocolumn,groupaddress]{revtex4-1}
\usepackage{graphicx}
\usepackage{amsmath,amssymb}
\usepackage{bm}
\usepackage{tikz}
\usetikzlibrary{positioning}

\begin{document}
\newcommand{\bra}[1]{\langle #1|}
\newcommand{\ket}[1]{|#1\rangle}
\newcommand{\braket}[2]{\langle #1|#2\rangle}
\newcommand{\HRule}[1]{\rule{\linewidth}{#1}}

\title{Proposal for an analog Schwarzschild black hole in condensates of light}
\author{L. Liao,$^{1}$ E. C. I. van der Wurff,$^{1}$ D. van Oosten,$^{2}$ and H.T.C. Stoof$^{1}$} 

\affiliation{$^1$Institute for Theoretical Physics and Center for Extreme Matter and Emergent Phenomena, Utrecht University, Princetonplein 5, 3584 CC Utrecht, The Netherlands\\
$^2$Debye Institute for Nanomaterials Science and Center for Extreme Matter and Emergent Phenomena, Utrecht University, Princetonplein 5, 3584 CC Utrecht, The Netherlands}

\date{\today}

\begin{abstract}
By etching a hole in the mirrors or by placing a scatterer in the center of a cavity, we can create a sink for light. In a Bose-Einstein condensate of photons this sink results in the creation of a so-called radial vortex, which is a two-dimensional analogue of a Schwarzschild black hole. We theoretically investigate the Hawking radiation and the associated greybody factor of this Schwarzschild black hole. In particular, we determine the density-density and velocity-velocity correlation functions of the Hawking radiation, which can be measured by observing the spatial correlations in the fluctuations in the light emitted by the cavity.
\end{abstract}

\maketitle

\section{Introduction}
Hawking found in 1974 that black holes radiate particles with a thermal spectrum \cite{Haw}, indicating that black holes have a Hawking temperature consistent with Bekenstein's earlier ideas on black-hole entropy and black-hole thermodynamics \cite{Bek73}. Calculating the black-hole entropy from a microscopic theory has ever since been an essential probe for candidates of quantum-gravity theories \cite{Strominger96}. The fact that the black-hole entropy is proportional to an area instead of a volume led to the formulation of the holographic principle \cite{Sus95} and to discussions regarding the black-hole information paradox \cite{Sus06}.

Due to the low Hawking temperatures of astronomical black holes, Hawking radiation has not been observed yet. To overcome this difficulty and to illuminate the conceptual issues of Hawking radiation, Unruh proposed a sonic analogue of a black-hole horizon \cite{Unr}. Such a sonic horizon is formed due to the incapability of sound waves to escape from a region where the fluid is moving faster than the speed of sound. Numerous theoretical proposals for analogue black-hole horizons in different systems \cite{Novello02,Balbinot13} are inspired by Unruh's work. These contains proposals for superfluid helium \cite{Jacobson98}, atomic Bose-Einstein condensates \cite{Garay00}, light in dispersive media \cite{Leonhardt00}, a fluid in a shallow basin \cite{Schutzhold2002}, electromagnetic waveguides \cite{Schutzhold05}, ultracold fermions
\cite{Giovanazzi05}, trapped-ion rings \cite{Horstmann10}, exciton-polariton condensates \cite{Solnyshkov11}, light in non-linear liquids \cite{Elazar12}, Weyl semi-metals \cite{Volovik16}, and magnons \cite{Jannes2011,Duine17}.

Experimental observations of different features of horizons have been made in water \cite{Rousseaux2008}, optical systems \cite{Philbin08,Faccio2010}, atomic Bose-Einstein condensates \cite{Steinhauer14}, and exciton-polariton condensates \cite{Nguyen15}. However, these experimental examples of analogue black-hole horizons are focussed on systems with a one-dimensional flow pattern that smoothly interpolates between a velocity below the speed of sound to a finite velocity above the speed of sound and thus do not contain a singular velocity profile. In this paper we propose to create in Bose-Einstein condensates of light \cite{Kla} the analogue of a rotationally symmetric Schwarzschild black hole. To do so requires a velocity singularity, which we implement by a radial vortex. The latter is a special, i.e., nonrotating, case of a draining bathtub vortex studied previously \cite{Basak2003,Berti}. As a result, our rotationally symmetric analogue sheds light on the properties close to the region of the singularity. Our sonic black-hole analogue also contains a horizon. Furthermore, the involved metric allows for an analytic solution of the phase fluctuations that incorporates the famous phase singularity at the horizon. Finally, we find a nontrivial greybody factor which modifies the thermal Bose spectrum of the Hawking radiation to a classical Maxwell-Boltzmann spectrum.

\begin{figure}
	\includegraphics[width=0.95\linewidth]{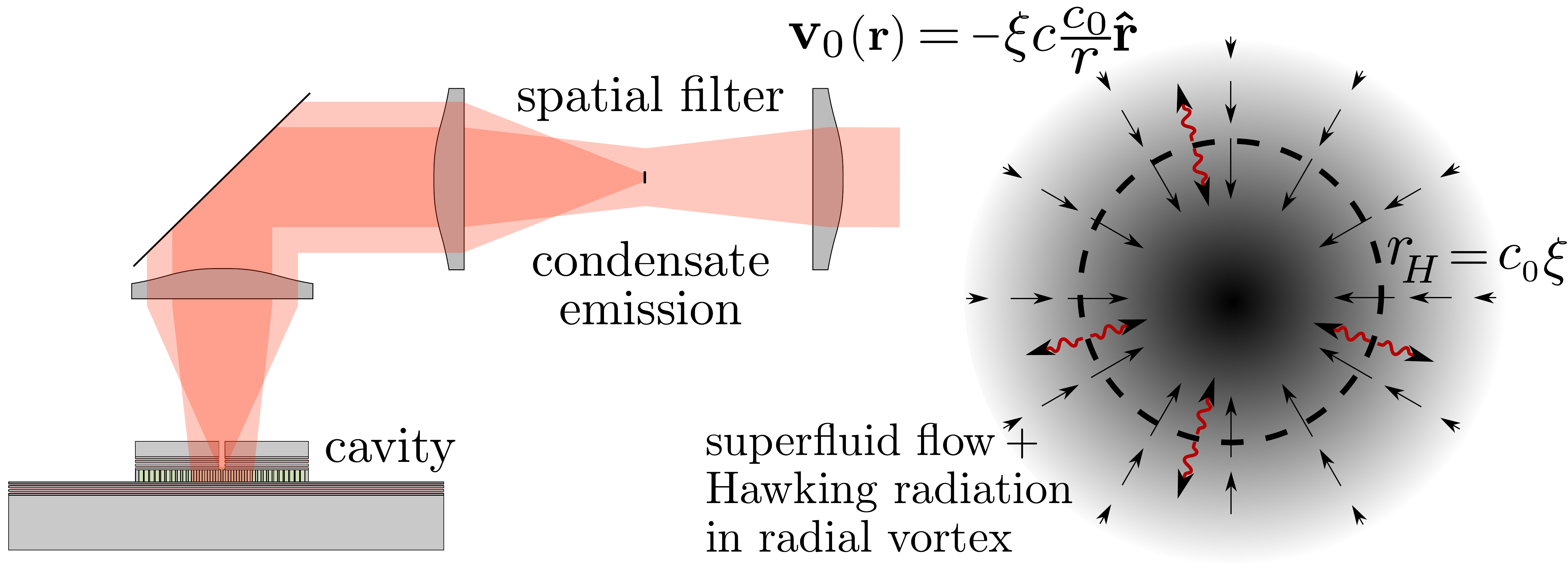}
	\caption{(Color online) We propose to etch a small hole in the highly reflecting mirrors of the cavity. Light from the condensate leaks out from this hole, which induces a radial flow velocity ${\bf v}_0({\bf r}) \propto - \hat{\bf r}/r$ in the photon condensate towards the hole. At some radius $r_H$ from the hole, the speed of this flow will exceed the speed of sound in the condensate. This results in a sonic horizon (indicated by the black dashed circle), from which Hawking radiation is emitted in the form of density fluctuations in the condensate. The light leaking from the hole will be blocked in an image plane. Light escaping vertically through the mirrors is sent through an additional lens, such that the angle of light emission is imaged. Hawking radiation in the condensate can then be observed by measuring spatial correlations in the noise of this emission.}\label{Fig4.1}
\end{figure}

To clarify our proposal we note from Fig.~\ref{Fig4.1} that in a radial vortex the superfluid velocity profile exactly resembles the electric field of a (negative) unit charge. In an atomic Bose-Einstein condensate such a radial vortex is not a solution of the superfluid hydrodynamic equations inside the core and therefore not physically allowed. The reason is that in this case particle number is conserved. As all velocities are directed radially inward, a sink of particles would be required in the core, which would quickly deplete the condensate. In the case of a photon condensate it is straightforward to create a sink for photons in the center of the trap by simply fabricating a hole in the cavity. Moreover, the loss of photons from the cavity can be compensated by pumping of the active medium in the cavity. The sink will thus not deplete the condensate and the radial vortex is a steady-state solution to the hydrodynamic equations of the photon condensate. We therefore propose to create such a radial vortex in a Bose-Einstein condensate of photons using an appropriately fabricated cavity.

The most important property of a radial vortex is that it is a Schwarzschild black-hole analogue and contains not only a horizon but also a singular behavior of the metric due to the singular velocity profile. Of course, a true metric singularity is ultimately cut-off by the vortex core, but we intend to study some of its effects by comparing systems with different sizes of the etched hole in the cavity. The horizon can be easily understood from the fact that the superfluid velocity strongly increases inversely proportional to the distance from the core, whereas the speed of sound for density fluctuations in the light fluid remains roughly constant as it is determined by the density of the photons. At a certain distance from the core these two velocities therefore become equal, which implies that fluctuations in the light intensity occurring closer to the core can no longer escape to the region outside the horizon as they will always be swept away into the sink in the core of the radial vortex.

As a consequence, the Schwarzschild black hole will emit Hawking radiation. Its origin is due to the fact that near the horizon the Bose-Einstein condensate of photons can create a pair of phonons, i.e., two quanta of density fluctuations, one moving through the horizon and ultimately falling into the sink in the vortex core, and one moving away from the horizon and escaping to large distances from the core where it contributes to the Hawking radiation. The possible reflection of the latter phonon as it moves through the light medium affects the frequency spectrum of the Hawking radiation and leads to deviations of the ideal black-body spectrum with the Hawking temperature. These deviations are described by the greybody factor \cite{Unr76,Schutzhold2002}, whose determination plays an important role in the quantum physics of Schwarzschild and other black-hole solutions in general relativity \cite{Das97}.

An important signature of the Hawking radiation is the spatial correlations between the fluctuations in the light intensity outside and inside the horizon. These correlations in intensity can be measured by imaging the light escaping from the cavity on a sensitive CCD camera and subsequently calculating autocorrelation functions of the fluctuations in those images. Specifically for this experiment, as illustrated in Fig.~1, we propose to collect the light from the cavity using an objective lens. Using a second lens, we create an intermediate image plane to block the light escaping from the leak in the cavity. The remaining light is thus light that leaked through the cavity mirror. Using a third lens, this light is collimated to image the angular distribution of the light leaking through the mirrors. Interestingly, such a correlation measurement can never be performed for a true astronomical black hole, where it is impossible to look behind the horizon. In our set-up this problem is avoided because the Bose-Einstein condensate is quasi two dimensional and the light needed for the correlation measurements escapes out of the plane of the cavity.

\section{Schwarzschild black hole in Bose-Einstein condensates}
We introduce the sound velocity $c\equiv \sqrt{gn_0/m}$ and the correlation (or healing) length $\xi\equiv\hbar/mc$, where $n_0$ is the condensate density, $m$ is the effective mass of the photon gas due to the cavity confinement and $g$ is the strength of the effective contact interaction. Then the radial-vortex has the following velocity profile
\begin{eqnarray}\label{eq4.1}
\textbf{v}_0(\textbf{r})=-\dfrac{\hbar c_0}{mr}\hat{\textbf{r}}=-\xi c \frac{c_0}{r}\hat{\textbf{r}},
\end{eqnarray}
with $c_0>0$ a constant, and $\hat{\textbf{r}}$ the unit vector in the radial direction. To show that this represents a Schwarzschild black-hole analogue in a two-dimensional Bose-Einstein condensate of light, we introduce dimensionless variables $r\rightarrow \xi r$ and $t\rightarrow \xi t/c$, and use these dimensionless variables in the following discussions.
From the hydrodynamical equations of a Bose-Einstein condensate \cite{Gr61,Pi61}, we obtain the equation of motion for the phase fluctuations
\begin{eqnarray}\label{eq4.2}
\left\{\nabla^2-\left(\partial_t-\dfrac{c_0}{r}\hat{\textbf{r}}\cdot\nabla\right)^2\right\}\delta\theta(\textbf{r},t)=0.
\end{eqnarray}
This can be rewritten as a wave equation in a curved spacetime
\begin{eqnarray}\label{eq4.3}
\nabla_\mu\nabla ^\mu\delta\theta(\textbf{r},t) \equiv
\frac{1}{\sqrt{-g}} \partial_\mu \sqrt{-g} g^{\mu\nu} \partial_{\nu} \delta\theta(\textbf{r},t) =0,
\end{eqnarray}
where the spacetime metric $g_{\mu\nu}$ in cylindrical coordinates $(r,\varphi)$ is given by \cite{Painleve1921}
\begin{eqnarray}\label{eq4.4}
ds^2=-\left(1-\frac{c_0^2}{r^2}\right)dt^2+2\frac{c_0}{r}drdt+dr^2+r^2d\varphi^2 .
\end{eqnarray}
In order to diagonalize the metric, we apply the coordinate transformation $d\tau=dt- c_0 dr/r(1-c_0^2/r^2)$ and the metric becomes
\begin{eqnarray}\label{eq4.5}
ds^2= -f(r)d\tau^2+f(r)^{-1}dr^2+r^2d\varphi^2,
\end{eqnarray}
with the warp factor $f(r)=1-c_0^2/r^2$. This warp factor is identical to that of a Schwarzschild black hole in $4+1$ spacetime dimensions with a mass parameter proportional to $c_0^2$. The horizon is at $r_H=c_0$ and the Hawking temperature is $T_H=df(r)/dr|_{r=r_H}/4\pi={1}/{2\pi c_0}$.

Note that the radial vortex exists in a spatially two-dimensional BEC and that we have now found that its effective metric equals the one of a Schwarzschild black hole in $4+1$ spacetime dimensions. This is a consequence of the fact that the metric of an acoustic black hole does not obey Einstein’s equations of General Relativity, but is determined by hydrodynamics. Indeed, Einstein’s equations do not even allow for a black-hole solution in $2+1$ spacetime dimensions in the absence of a cosmological constant. The most important characterizations of the metric, and hence the name of the analogue black hole, are therefore the rotational symmetry and the form of the warp factor, that describes the spacetime curvature in the radial and time directions.

Quantization of the phase fluctuations is achieved by expanding in a basis of eigenfunctions $\psi_m(\omega,r)e^{im\varphi-i\omega\tau}$ of the wave equation in Eq.\,\eqref{eq4.3}, i.e.,
\begin{eqnarray}\label{eq4.6}
\delta\hat{\theta}(r,\varphi,\tau)&&=
\int_{0}^{\infty}d\omega \sum_{m=-\infty}^{\infty}\left\{e^{-i\omega\tau}e^{im\varphi}\psi_{m}(r,\omega) \hat{a}_{mR}(\omega) \right. \nonumber\\
&&\left. + e^{-i\omega\tau}e^{im\varphi}\psi^*_{m}(r,\omega) \hat{a}_{mL}(\omega)+ \mathrm{c.c.}\right\},
\end{eqnarray}
where we defined for each angular momentum $m$ the right-moving and left-moving operators, $\hat{a}_{mR}(\omega)$ and $\hat{a}_{mL}(\omega)$ respectively, which satisfy the commutation relations $[\hat{a}_{mR(L)}(\omega),\hat{a}^{\dagger}_{m'R(L)}(\omega')]_-=\delta(\omega-\omega')\delta_{mm'}$.

\section{Hawking radiation and greybody factors}
\begin{figure}[t]
	\includegraphics[scale=0.3]{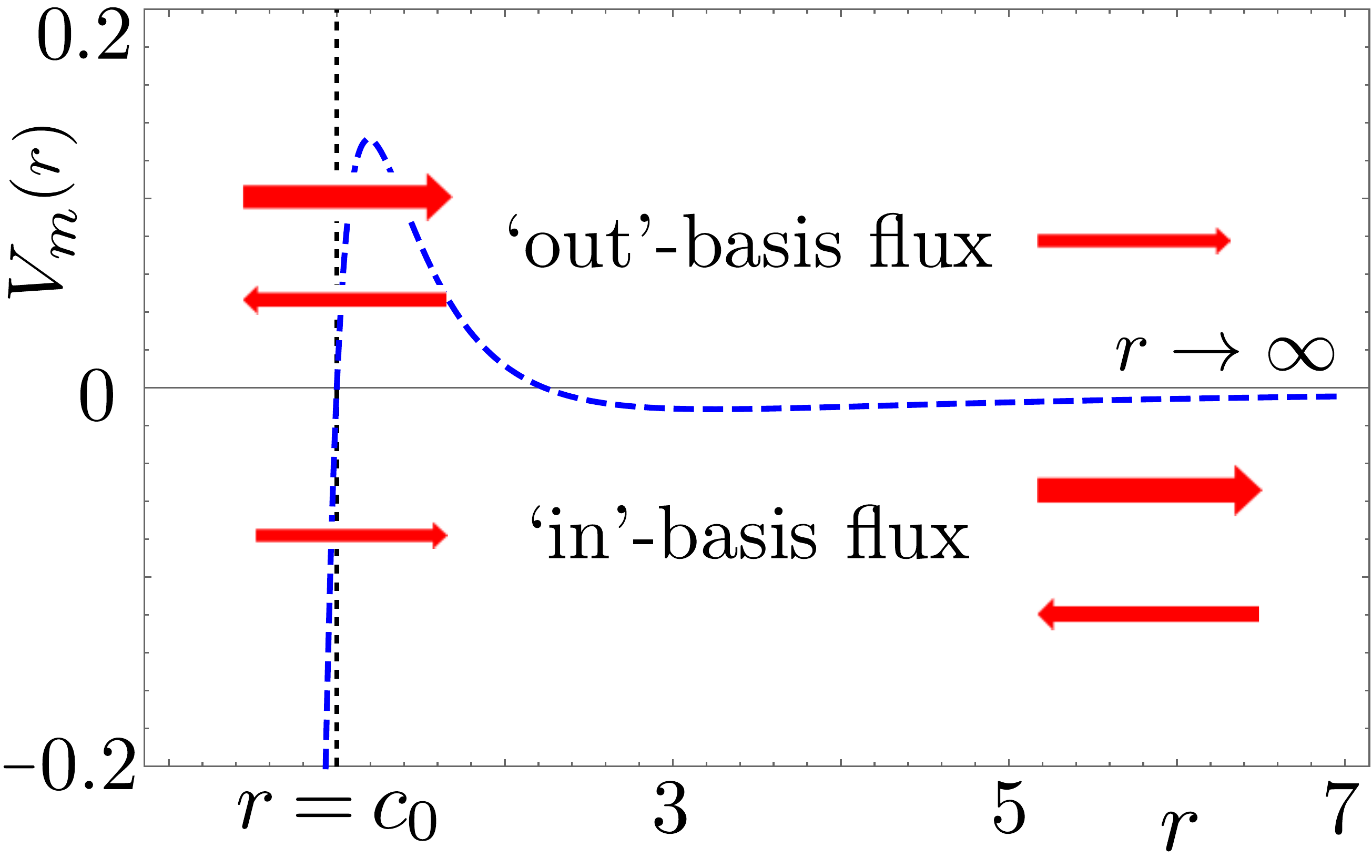}
	\caption{(Color online) The flux composition of the `out' and `in' basis wave functions of the analog Hawking effect. The `out' basis wave functions contain only outgoing flux at $r\rightarrow\infty$, whereas the `in' basis wave functions contains only outgoing flux at the horizon located at the dimensionless radius $r=c_0$. The blue dashed curve illustrates the effective potential $V_m(r)=(1-{c_0^2}/{r^2})({5c_0^2}/{4r^2}+m^2-{1}/{4})/{r^2}$ for $c_0=1$ and $m=0$. Note the tunneling barrier in this potential just outside the black hole, which causes the sharp reduction of the greybody factors at low frequencies shown in Fig.~\ref{Fig4.3}. The dotted, vertical black line indicates the location of the horizon.}\label{Fig4.2}
\end{figure}

In order to discuss the Hawking radiation in our set-up, we define two bases of wave functions. The `out' basis is defined by requiring only out-going waves as the radius approaches infinity, and the `in' basis is defined by having only out-going waves at the horizon. Fig.~\ref{Fig4.2} illustrates the flux composition of the `out' and `in' basis wave functions. In what follows, the `in' basis is needed to define the vacuum state of the black hole that contains no infalling phonons at infinity, whereas the `out' basis measures the outgoing flux of phonons in that state, which constitutes the Hawking radiation.

For angular momentum $m=0$, the analytical solution of Eq.~\eqref{eq4.3} is
\begin{subequations}
	\begin{align}
	\psi_{0}(r,\omega)&={\cal N}\bigg\{c_1I_{-ic_0\omega}{(i\omega\sqrt{r^2-c_0^2})}. \nonumber\\
	&\phantom{=} +c_2 I_{ic_0\omega}{(i\omega\sqrt{r^2-c_0^2})}\bigg\} \quad {\rm for~} r>c_0, \label{eq4.7a} \\
	\psi_{0}(r,\omega)&={\cal N} e^{-\pi c_0\omega/2} \bigg\{c_1
	e^{\pi c_0\omega}I_{-ic_0\omega}{(\omega\sqrt{c_0^2-r^2})} \nonumber\\
	&\phantom{=} +c_2 I_{ic_0\omega}{(\omega\sqrt{c_0^2-r^2})}\bigg\}, \quad {\rm for~} r<c_0 	 \label{eq4.7b},
	\end{align}
\end{subequations}
where the normalization ${\cal N}={1}/{\sqrt{4\pi n_0}(e^{2\pi c_0\omega}-1)}$ follows from the commutation relations of the phase and density operators, and  $I_\alpha(z)$ is the modified Bessel function. Note that near the horizon $I_{\mp ic_0\omega}{(\omega\sqrt{c_0^2-r^2})}\propto e^{\mp i\omega c_0 \ln(r-c_0)/2}$, which incorporates the well-known phase singularity that is the trademark of a horizon. The solution explicitly satisfies the continuity equation for the photon density at the horizon. For the `out' basis wave function $\psi_{0}^{\rm out}(r,\omega)$, the coefficients are $c_1^{\rm out}=-1$, $c_2^{\rm out}=e^{2\pi c_0\omega}$. And for the `in' basis wave function $\psi_{0}^{\rm in}(r,\omega)$, $c_1^{\rm in}=0$, $c_2^{\rm in}=e^{\pi c_0\omega}\sqrt{e^{2\pi c_0\omega}-1}$.

By investigating the Bogoliubov transformation from the `in' basis to the `out' basis, we can obtain the Bogoliubov coefficients and the Hawking radiation. We find that
$\psi_{0}^{\rm out}(r,\omega)=u_{0}(\omega)\psi_{0}^{\rm in}(r,\omega)
-v_{0}(\omega){\psi_{0}^{\rm in}}^*(r,\omega)$,
with $|v_{0}(\omega)|^2\equiv |u_{0}(\omega)|^2-1=1/(e^{2\pi c_0\omega}-1)$. Hence, the Bogoliubov transformation for $m=0$ is given by
\begin{eqnarray}\label{eq4.8}
\hat{a}_{0R(L)}^{\rm out}(\omega) =u_{0}(\omega)\hat{a}_{0R(L)}^{\rm in}(\omega) +v_{0}(\omega)\hat{a}_{0R(L)}^{\rm in \dagger} (\omega).
\end{eqnarray}
Defining the `in' vacuum state by the requirement $\hat{a}_{mR(L)}^{\rm in}|0\rangle_{\rm in}=0$, we thus have
$_{\rm in}\langle0|\hat{a}_{mR(L)}^{\rm out \dagger}\hat{a}_{mR(L)}^{\rm out}|0\rangle_{\rm in} =|v_{0}(\omega)|^2=1/(e^{\omega/T_H}-1)$,
which is the Hawking effect.

\begin{figure}[t]
	\begin{tikzpicture}
	\node (img)  {\includegraphics[scale=0.35]{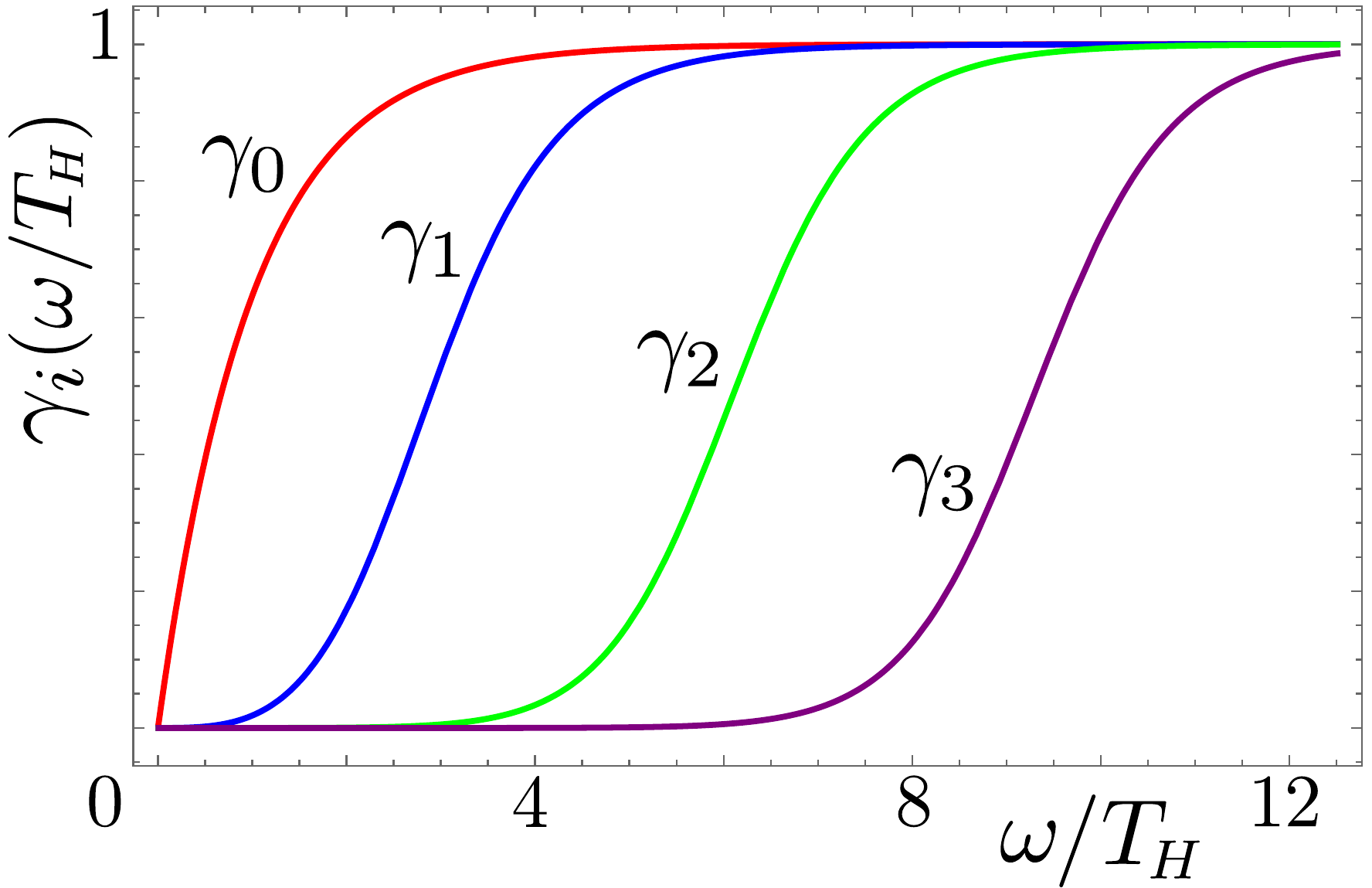}};
	\end{tikzpicture}
	\caption{(Color online) The (from left to right) red, blue, green, and purple curves show the greybody factor $\gamma_m$ as a function of $\omega/T_H$ for $|m|=0, 1, 2, 3$, respectively.}\label{Fig4.3}
\end{figure}

Another interesting aspect associated to the Hawking radiation is the greybody factor, which can be understood by introducing the tortoise coordinate
\begin{equation}
r^* = r - \frac{c_0}{2}\ln\bigg|\frac{r+c_0}{r-c_0}\bigg|,
\end{equation}
to transform the wave equation to a Schr\"{o}dinger equation with an effective potential $V_m(r^*(r))$ as shown in Fig.~\ref{Fig4.2} \cite{Schutzhold2002,Berti}. The wave propagating from the horizon to infinity is scattered by the effective potential outside the horizon. For $m=0$, we obtain the greybody factor
$\gamma_0(\omega)=1-e^{-\omega/T_H}$
analytically from the $S^{\rm out}$-matrix of Eq.~(\ref{eq4.7a}),
which is given by the ratio between the out-going flux at infinity and at the horizon. In general, the total number of emitted particles is
\begin{eqnarray}\label{eq4.9}
N=\int_{0}^{\infty} d\omega \frac{\omega}{2\pi}  \frac{\sum_m \gamma_m(\omega)}{e^{{\omega}/{T_H}}-1}
\equiv \int_{0}^{\infty} d\omega \frac{\omega}{2\pi}  \rho_H(\omega)~,
\end{eqnarray}
where for $m\neq 0$ we determine the greybody factors numerically as shown in Fig.~\ref{Fig4.3}. Note that at low frequencies the spectrum is dominated by $m=0$ and $\rho_H(\omega) \simeq e^{-\omega/T_H}$, so the quantum Bose spectrum has changed into the classical Maxwell-Boltzmann spectrum. This is an interesting effect of the nontrivial effective potential resulting from the curvature of the spacetime.

\section{Correlation functions}
To show that the Hawking radiation is a result from entangled phonon pairs created near the horizon \cite{Haw}, we need to study the correlation functions of the light fluctuations in the Bose-Einstein condensate \cite{Balbinot13,Busch14,Steinhauer15,Rousseaux2016}. We therefore define the conserved photon current density
$\hat{J}^\mu({\bf r},\tau)\equiv g^{\mu\nu}\partial_\nu\delta\hat{\theta}({\bf r},\tau)=\nabla^\mu \delta\hat{\theta}({\bf r},\tau)$
and obtain for the current-current correlation function
\begin{eqnarray}\label{eq4.10}
\Pi&&^{\mu\nu}({\bf r},{\bf r}';\tau-\tau') \equiv
~_{\rm in}\langle0|\hat{J}^\mu({\bf r},\tau)\hat{J}^\nu({\bf r}',\tau')|0\rangle_{\rm in} \nonumber\\
= &&\int_{0}^{\infty}d\omega \sum_{m=-\infty}^{\infty} \chi_c(\omega) \left(\frac{1}{e^{\omega/T_H}-1}+\frac{1}{2}\right)\times \label{eq:currentcurrent}\\
&&\nabla^\mu \{\psi^{\rm out}_{m}(r,\omega)e^{im\varphi-i\omega \tau}\}
\nabla^\nu \{{\psi^{\rm out}_{m}}^*(r',\omega)e^{-im\varphi'+i\omega \tau'}\},\nonumber
\end{eqnarray}
which due to Eq.~\eqref{eq4.3} satisfies the desired Ward identity
$\nabla_\mu \Pi^{\mu\nu}({\bf r},{\bf r}';\tau-\tau')=0$ for all ultra-violet cut-off functions $\chi_c(\omega)$.

\begin{figure}[h!]
	\begin{tikzpicture}
	\node (img)  {
		\includegraphics[width=0.75\linewidth]{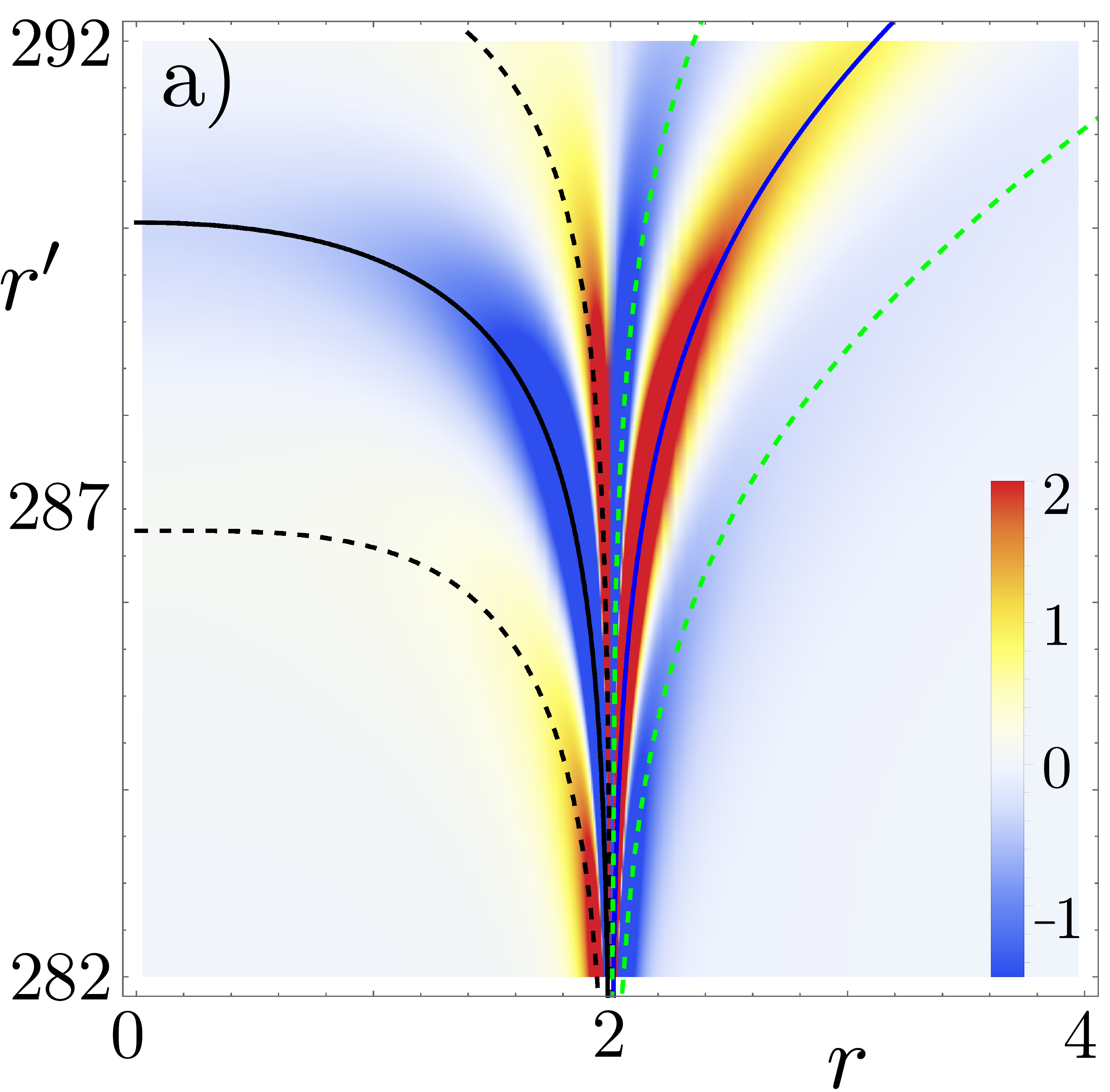}
	};
	\end{tikzpicture}
	\begin{tikzpicture}
	\node (img)  {
		\includegraphics[width=0.75\linewidth]{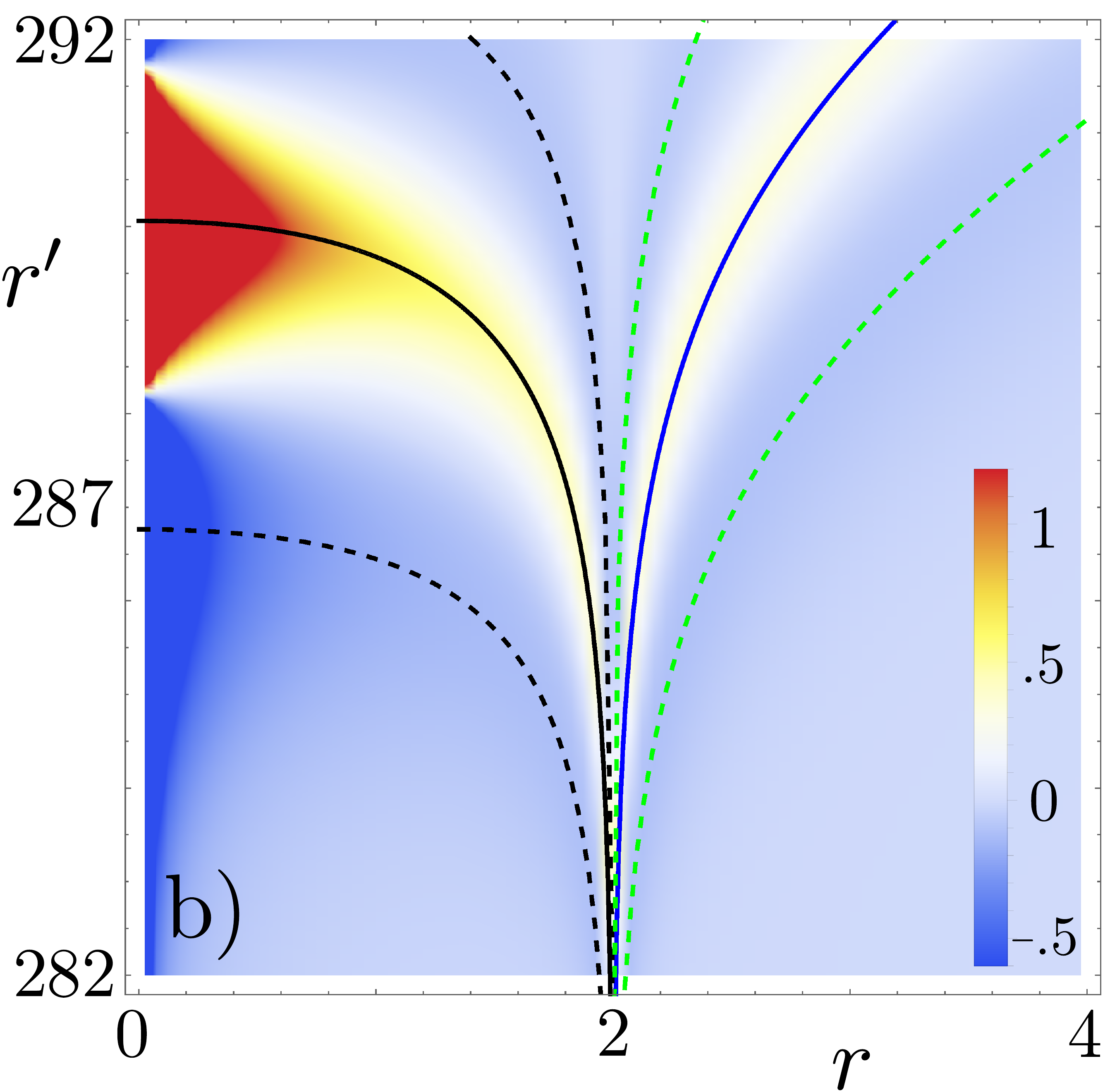}
	};
	\end{tikzpicture}
	\caption{(Color online) Color plot of a) the quantum part of the density-density correlation function $\Pi^{tt}(r,r';\Delta t_c)\times 10^{3}$ and b) the quantum part of the velocity-velocity correlation function $\Pi^{rr}(r,r';\Delta t_c)$, both for angular momentum $m=0$ and $c_0=2$. We chose $\Delta t_c=-300$, which is sufficiently large to be in the hydrodynamic regime. The black solid curve (second line from the left) corresponds to the expected peak position in the ray approximation for the dimensionless radii $r<c_0$ and $r'>c_0$. The blue solid curve (second line from the right) corresponds to the expected peak position in the ray approximation for $r>c_0$ and $r'>c_0$. The black dashed curves (first and third line from the left) and green dashed curves (first and third line from the right) show the secondary peaks of the correlation functions, which have the opposite sign of the corresponding primary peaks. In a) the left (right) primary peak is negative (positive), whereas in b) both primary peaks are positive. Note that the density-density correlation function is singular near the horizon, but in such a narrow region that it is not visible here.}\label{Fig4.4}
\end{figure}

In particular, $\Pi^{\tau\tau}({\bf r},{\bf r}';\tau-\tau')$ is the density-density correlation function and $\Pi^{rr}({\bf r},{\bf r}';\tau-\tau')$ is the (radial) velocity-velocity correlation function. As can be seen explicitly in Eq.\eqref{eq4.10}, the correlation functions contain a thermal-like part that is proportional to the Bose-Einstein distribution function. It turns out that this contribution is negligible for $c_0= {\cal O}(1)$. In the following, we therefore focus on the remaining, ``quantum'' part of the correlation functions. Furthermore, we note that to determine the density-density and velocity-velocity correlation functions that can be measured in an experiment, we need to perform a coordinate transformation from $\tau$ back to the physical time $t$. Our final results for the angular-averaged quantum parts are obtained in this manner and are shown in Fig.~\ref{Fig4.4}.
Our quantum theory for the phase fluctuations is valid in the hydrodynamic regime, which in our dimensionless units implies $\omega \ll 1$. To remain in this regime we should consider the long-wavelength limit of the correlation functions, i.e., $|\Delta\tau|/2\pi \equiv |\tau - \tau'|/2\pi \gg 1$. Doing so indeed makes the correlation functions essentially independent of the cut-off function $\chi_c(\omega)$ in Eq.~\eqref{eq4.10}, except near the dominant correlation lines in the $(r,r')$ plane, whose locations are however universal as we show shortly. In our results we have used the cut-off function $\chi_c(\omega)=1/(1+\omega^4)$ that is appropriate for a point interaction between the photons \cite{Pe08,Henk}, but this may need reconsideration once the details of the gain material in the experimental set-up are better known.

The location of the dominant correlation lines can be easily understood in the ray approximation for phonons moving in the spacetime metric in Eq.~\eqref{eq4.4}. Requiring $ds^2=0$, we obtain $dt=dr/(1-c_0/r)$.
Consequently, the phonons of a Hawking pair emitted from the horizon arrive at the positions $r$ and $r'$ with a  time difference equal to $\Delta t(r)-\Delta t(r')$, where $\Delta t(r) \equiv r+ c_0 \ln|r/c_0-1|$. For a time delay $\Delta t_c$ in the correlation function the expected location of the dominant correlation line is therefore determined by $\Delta t(r)-\Delta t(r')=\Delta t_c$, which turns out to be in excellent agreement with our numerical results in Fig.~\ref{Fig4.4}.. The secondary correlation lines are due to the wave nature of the phonons and are caused by an overshoot in the destructive interference of the wave functions $\psi^{\rm out}_0(r,\omega)$ in the integral over frequencies involved in the correlation functions.

\section{Discussion and outlook}
We have discussed the properties of a radial vortex in a Bose-Einstein condensate of light, which is a Schwarzschild black-hole analogue. The Hawking radiation emitted by this analogue black hole consists of entangled phonon pairs in the light fluid that may have interesting applications for new light sources with unusual correlation properties.

Throughout this paper we have assumed that the hydrodynamic regime holds, which allowed us to write down the effective metric in Eq.\,\eqref{eq4.5}. At short distances from the horizon, this approximation may break down and so-called trans-planckian physics may start playing an important role. This is signaled by the fact that for a nonzero frequency $\omega$ the wave functions in Eqs.\,\eqref{eq4.7a} and \eqref{eq4.7b} oscillate ever more rapidly as the horizon is approached. The inclusion of trans-planckian physics ensures regularity of the solution by effectively providing an ultraviolet momentum cut-off at the horizon. For our particular case, however, this does not affect our results significantly, provided that we work at an energy $\omega c_0 \ll 1$. One way to understand this is that for $\omega = 0$ there is no phase singularity and trans-planckian effects are absent. Physically, this implies that this is also the case for $\omega c_0 \ll 1$, since the wave functions depend on $\omega c_0$. More mathematically, we can perform a transformation of the wave function such that the wave equation becomes a Schr\"odinger equation with an effective potential $V_{\text{eff}}(r)$ that diverges as  $-(1+(\omega c_0)^2)/4(r-c_0)^2$ near the horizon. Due to trans-planckian physics the divergence proportional to $(\omega c_0)^2$ is cut off. However, when $\omega c_0 \ll 1$ the resulting change in the effective potential does not lead to any backscattering and so does not affect the coefficients of the Bogoliubov transformation that determine the Hawking spectrum and the greybody factors. As we are especially interested in the regime $\omega c_0 \lesssim c_0 T_H = 1/2\pi$ the above condition is fulfilled and we indeed do not expect any significant corrections due to trans-planckian physics. (See for instance also Ref.~\cite{Planck} that comes to the same conclusion by a different analysis.) Note that this argument would not apply for a system that is described by an effective potential with a frequency-dependent dominant divergence.

In the future we also want to explore current-current correlation functions that are not angularly averaged, because of recent proposals of angular correlations in the Hawking radiation emitted at angle $\varphi$ and $\varphi+\pi$ due to the presence of a singularity in the center of the black hole \cite{Gerard}. This is beyond the scope of the present paper where we have focused on analytical results that most clearly demonstrate the physics of the radial vortex.

Finally, we believe that Bose-Einstein condensates of light in a semiconductor microcavity may be particularly promising experimentally \cite{arie-willem}. The existing dye-based set-ups \cite{Kla} appear not to have sufficiently strong interactions \cite{Wurff}. That is, to be in the hydrodynamic regime, where the Bose-Einstein condensate of photons allows for a fluid description and all relevant energy and length scales for our proposal are set by the chemical potential $gn_0$, the chemical potential needs to be much larger than the energy splitting of the external trap. Experimentally, the external trap is due to the curvature of the cavity mirrors. Hence, instead of trying to tune the interaction strength, it may be more convenient to simply reduce the energy splitting by using cavity mirrors with a smaller curvature. Note that the above condition for the validity of the hydrodynamic approximation also makes sure that the radius of the black-hole horizon $r_H = c_0$ is much smaller than the size of the light condensate, as is clearly required.

\begin{acknowledgements}
We thank Govert Nijs, Nick Plantz, and Stefan Vandoren for helpful discussions. This work is supported by the China Scholarship Council (CSC), the Stichting voor Fundamenteel Onderzoek der Materie (FOM) and is part of the D-ITP consortium, a program of the Netherlands Organization for Scientific Research (NWO) that is funded by the Dutch Ministry of Education, Culture and Science (OCW).
\end{acknowledgements}

\end{document}